# Attraction to natural stimuli in *Drosophila melanogaster*

Agnieszka Ruebenbauer[*]

Chemical Ecology-Ecotoxicology

Department of Ecology, Lund University

SE-223 62 Solvegatan 37, Lund, Sweden

Electronic address: Agnieszka.Ruebenbauer@yahoo.pl

[*]*Corresponding author*

**Abstract**

The animals, in particular insects (*Drosophila melanogaster*), response towards odor stimuli in nature can be established by measuring the dynamic of the odor response. Such an approach is innovative since responses to odors were tested only at certain time point so far, not in hour intervals for several hours. The odor attraction to 14 natural and ecologically relevant odor stimuli such as: fruits- ripe and rotten, yeast and vinegar was tested. A mathematical model to evaluate obtained data is proposed in this study in which the number of flies caught over several time points is presented as one simple parameter showing trapping potential of the trap housing particular odor stimuli. The knowledge concerning dynamic of the odor response in *Drosophila melanogaster* may enlighten the principles of flies behavior in context of exposure towards odor stimulus.

# Introduction

Host plant volatiles that attract insect herbivores by offering good oviposition sites are of crucial interest with respect to insect ecology and evolution. A better knowledge of insects-plant interaction helps in development of novel crop protection strategies. Identification of behaviorally attractive plants or plant components and defining processes underlying host choice detection may lead to better understanding chemical detection in insects and become significant for plant breeding for insects resistance.

Simply, fruit odor recognition by insects can serve as a model system for understanding how animals process complex environmental information (Budick and Dickinson, 2006; Jefferis et al., 2007). Each fruit originating from each plant species produces unique scent, containing complex blend of low molecular compounds (Jordan et al., 2002). Different fruit species usually share many volatile components (Herrmann, 1995), however their concentration and combination remains species specific (do Nascimento et al., 2008). Moreover the same fruit may serve as perfect oviposition site only at certain stage of ripeness. The elegant example of such a situation is fruiting part of fig changing its volatile composition over time and ovipositing on it *Drosophila* species (Lachaise et al.,1982).

There were reviewed several experiments conducted to understand the intricate principles of insects odor recognition on the level of odor oriented behavior (Carde and Willis, 2008; Carlson, 1996; Devaud, 2003), however there is little focus on models trying to explain the process of odor detection.

In the current study a model describing the two choice experiment is described where tested stimulus versus control are presented to the fly (first similar chemoperception two choice test was proposed by Tanimura et al. (1982) as a feeding preference test and modified later on by others). In the presented model flies kept in the experimental chamber can, based on their odor preferences, freely make choice over elapsed time. The odor driven behavior stated by flies entering the traps was recorded for 8h in hour intervals and after 24h. The ecologically relevant odor stimuli, 6 fruits at ripe and rotten stage, yeast and balsamic vinegar were put into traps and opposed to the control. In the presented model it is shown how to express the odor attraction in a clear and simple way, how to compress the obtained behavioral results- number of caught flies into the trap over elapsed time.

**Basic principles of the model used to evaluate data**

Let us consider some closed space filled at the local time instant $t = 0$ with $N_0 > 0$ flies. The flies could be trapped permanently by one of the two traps embedded into the closed space. A trapping potential for the above traps depends neither on time nor on the number of flies already trapped by the particular trap. Let us assume that flies do not interact between themselves and they do not have memory of the past events. Under such circumstances the number of flies remaining out of the traps is described by the following differential equation for $t \geq 0$:

$$\frac{dN}{N} = -(\lambda_1 + \lambda_2)\, dt.$$

(1)

Here the symbol $dN$ stands for the infinitesimal number of flies removed by traps at given time. The symbol $N = N(t)$ stands for the number of flies remaining out of traps at the same time $t$, constant parameters $\lambda_1 > 0$ and $\lambda_2 \geq 0$ describe trapping potential of respective traps, while the symbol $dt > 0$ denotes infinitesimal increment of the local time into the future. The number $dN$ is negative as traps remove flies. Usually the second trap described by $\lambda_2 \geq 0$ is considered as the reference trap in particular experiment. The main task is to estimate trap parameters $\lambda_1$ and $\lambda_2$ by using experimental data at hand. Equation (1) could be integrated using predefined initial condition, and one obtains the following solution for the present case:

$$N(t) = N_0 \exp[-(\lambda_1 + \lambda_2)t].$$

(2)

On the other hand, the number of flies already trapped is the following function of time:

$$n(t) = N_0 - N(t) = N_0 \left(1 - \exp[-(\lambda_1 + \lambda_2)t]\right).$$

(3)

The last equation (3) is valid due to the fact that the total number of flies $N_0$ is constant. Trapped flies are divided between two accessible traps according to the expression:

$$n(t) = n_1(t) + n_2(t) \text{ with}$$

$$n_1(t) = N_0 \left( \frac{\lambda_1}{\lambda_1 + \lambda_2} \right) (1 - \exp[-(\lambda_1 + \lambda_2)t]) \text{ and}$$

$$n_2(t) = N_0 \left( \frac{\lambda_2}{\lambda_1 + \lambda_2} \right) (1 - \exp[-(\lambda_1 + \lambda_2)t]).$$

(4)

Here the index 1 refers to the first trap, while the index 2 refers to the second trap, respectively. All of the above functions of time are directly accessible experimentally. On the other hand, one can define the following dependent function of time:

$$y(t) = \frac{n_1(t) - n_2(t)}{N_0} = \left( \frac{\lambda_1 - \lambda_2}{\lambda_1 + \lambda_2} \right) (1 - \exp[-(\lambda_1 + \lambda_2)t]).$$

(5)

Subsequently one can introduce two new parameters dependent on the trapping potentials, i.e., $\lambda = \lambda_1 + \lambda_2 > 0$ and $\beta = (\lambda_1 - \lambda_2)/(\lambda_1 + \lambda_2)$. For this new representation equation (5) takes on the form:

$$y(t) = \beta (1 - \exp[-(\lambda t)]).$$

(6)

One can define another dependent function of time as:

$$x(t) = -\ln\left[ \frac{N(t)}{N_0} \right] = \lambda t.$$

(7)

The last function described by equation (7) is a straight line going through the origin and having positive slope. Hence, simple one-parameter linear regression could be used to determine parameter $\lambda$. Hence, one can calculate next dependent function of time:

$$z(t) = 1 - \exp[-(\lambda t)] \text{ with } 0 \leq z(t) < 1.$$

(8)

The last function (8) could be calculated for any desired time. Finally, it could be observed that the following relationship $y(t) = \beta z(t)$ holds. Therefore one can apply again simple one-parameter linear regression fit through the origin in order to obtain parameter $\beta$. Upon having above parameters one can transform back to the original parameters applying the following transformations:

$$\lambda_1 = \left(\frac{\lambda}{2}\right)(1+\beta) \text{ and } \lambda_2 = \left(\frac{\lambda}{2}\right)(1-\beta).$$

(9)

One has to note that the total number of flies $N_0$ is known *a priori* and therefore it is not an adjustable parameter. Observations should be made for definitely defined positive and increasing time instants different one from another. Usually errors of the time scale are negligible. It is sufficient to measure functions $n_1(t)$ and $n_2(t)$ as $N(t) = N_0 - [n_1(t) + n_2(t)]$. Essentially it is two-parameter data fit, and therefore observations have to be made at three distinctly different time instants at least. Usually the second trap has neutral potential. Hence, one has attractive potential of the first trap provided $\lambda_1 > \lambda_2$ and repulsive potential in the opposite case, i.e., for $\lambda_1 < \lambda_2$. A special case $\lambda_1 = \lambda_2$ is inaccessible due to the experimental errors.

There is no reference for $\lambda_2 = 0$ as one cannot observe repulsive potential of the first trap in such case. It means that one has too small cross-section of traps in comparison with the number of objects injected $N_0$. Cross-section of the trap is the measure of the statistical attraction, i.e., it adds to the real potential. Both traps have to have the same cross-sections in order to obtain reliable results. Errors of the parameters are to be estimated taking into account deviations from the fitted straight line. For the first fit [equation (7)] errors on the abscissa are usually negligible. The situation is more complex for the second fit. Once the error of the parameter $\lambda$ is found from the first linear regression fit as $\delta\lambda$ one can calculate error of the abscissa for the second fit as $\delta z = t \exp[-(\lambda t)]\,\delta\lambda$. Errors of the parameters described by equation (9) could be estimated as:

$$\delta\lambda_1 = \tfrac{1}{2}\sqrt{\delta\lambda^2\,(1+\beta)^2 + \lambda^2\,\delta\beta^2} \quad \text{and} \quad \delta\lambda_2 = \tfrac{1}{2}\sqrt{\delta\lambda^2\,(1-\beta)^2 + \lambda^2\,\delta\beta^2}.$$

(10)

Here the symbol $\delta\beta$ denotes error of the parameter $\beta$. The role of traps could be easy exchanged, of course, and hence the choice of the reference trap is arbitrary as far as one considers above model. In order to perform effectively transformation described by equation (7) one needs to fulfill the following condition $N(t) > 0$. In the absence of the first trap one obtains $\beta = -1$, $\lambda_1 = 0$ and $\lambda_2 = \lambda$, i.e., some trivial extension of the above model. In the absence of the second trap one has $\beta = 1$, $\lambda_1 = \lambda$ and $\lambda_2 = 0$. One can repeat experiments under similar conditions, of course, and calculate respective average with the corresponding error. The most important parameter is $\lambda_1 - \lambda_2$ or the average $\langle \lambda_1 - \lambda_2 \rangle$ provided the experiment has been repeated several times. Weighed average $\langle \lambda_1 - \lambda_2 \rangle_W$ could be calculated as well. In addition effective dispersions of above parameters could be calculated provided the number of repetitions is greater than unity. The last parameters are positive for attracting traps and negative for repulsive traps. Neutral traps should yield above parameters close to zero. Example of attractive trap is shown in Figure 1. This is example of strongly attractive trap as for $N_0 = 29$ one obtains $\lambda_2 = 0$. Hence, the sole relevant function is $n_1(t)$. Statistical errors due to the small number of objects $N_0$ prevent further complication of the model used.

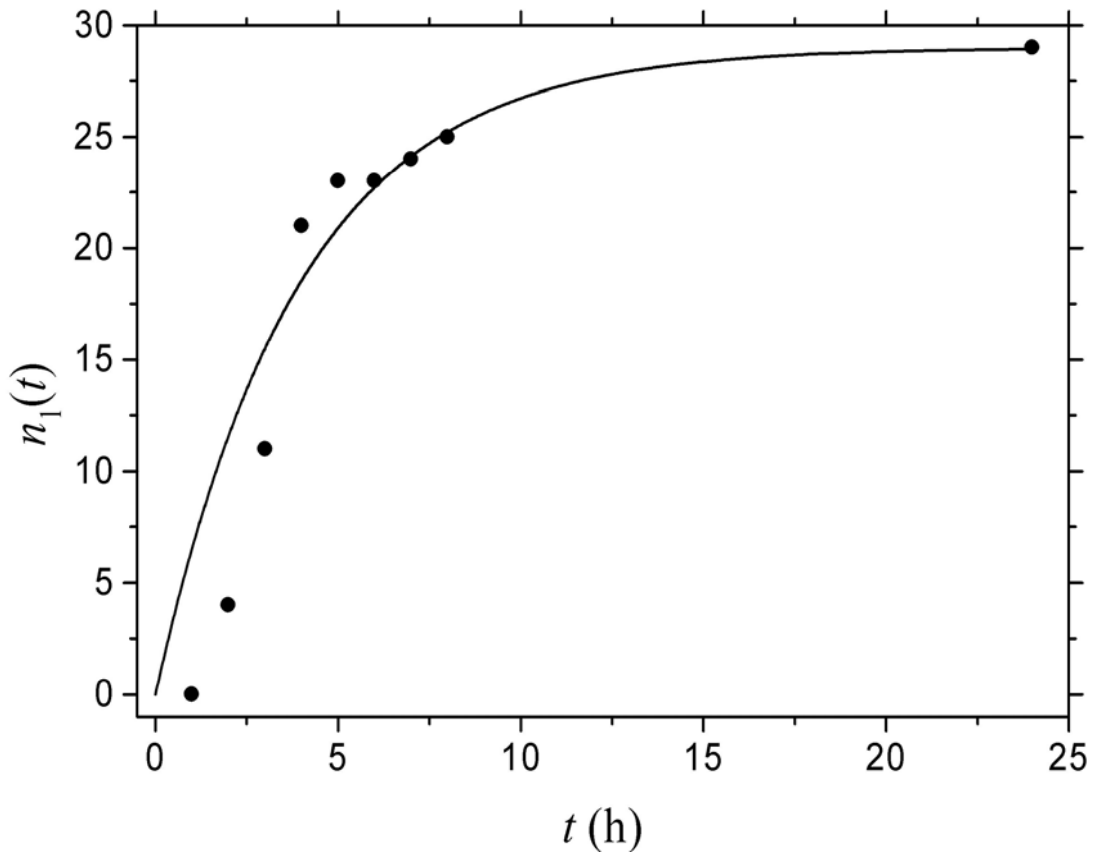

**Figure 1** Function $n_1(t)$ obtained for line CS with trap 1 filled with apple: run No1 with $N_0 = 29$. All objects have been trapped by trap 1 after sufficiently long time as shown.

Model described above is a special case of the more general model described by set of the following differential equations:

$$\frac{dn_k}{dt} = \left[ N_0 - \sum_{l=1}^{K} n_l(t) \right] \lambda_k(t) \text{ with}: t \geq 0, n_k(0) = 0, N_0 > 0, K \geq 1 \text{ and } \lambda_k(t) \geq 0.$$

(11)

Solutions of the above differential equations take on the following form:

$$n_k(t) = \int_0^t dt' \left[ N_0 - \sum_{l=1}^{K} n_l(t') \right] \lambda_k(t').$$

(12)

Functions $\lambda_k(t)$ of the local time $t$ are to be set in accordance with the particular dynamics of the system. Extended model describes $K$ final states. The situation is more complicated for imperfect traps, i.e., traps being able to release trapped objects. Such situations should be avoided unless this effect is really minor.

Parameter $\lambda$ describing attraction or repulsion could be expressed in the following way for isotropic random gas composed of non-interacting diluted objects provided attraction or repulsion is weak:

$$\lambda = \sigma \sqrt{\frac{\langle v^2 \rangle}{3V^2}}.$$

(13)

Here the symbol $\sigma \geq 0$ stands for the effective cross-section of the trap, the symbol $\langle v^2 \rangle \geq 0$ denotes mean squared velocity within the ensemble of flies remaining out of traps, and finally, the symbol $V > 0$ stands for the volume of the closed space accessible to the objects remaining out of traps. It is assumed that the space has three dimensions and remains Euclidean. The mean squared velocity is referred to the immobile traps. All of the above quantities remain practically constant versus local time for properly designed trap. An exception could be mean squared velocity provided flies are complex living beings.

**Material and methods**

All flies used in the conducted experiments, ecology relevant natural stimuli and performed trap assay were as in supplementary material, Ruebenbauer et al. (2008). A detailed description of traps used was modified after Park et al. (2002), Zhu et al. (2003) and Dekker et al. (2006) and could be found in Ruebenbauer et al. (2008) supplementary material. Each experiment has been repeated five times with the neutral trap filled either with distilled water or non-volatile paraffin.

**Results and discussion**

14 ecology relevant natural stimuli, such as 5 fruits at ripe or rotten stage, yeast and vinegar were tested on 5 different *Drosophila melanogaster* strains under semi natural conditions. The origin and history of the tested strains is reported in supplemental material in Ruebenbauer et al. (2008). The results are shown in fig. 2 where parameter $\langle \lambda_1 - \lambda_2 \rangle_W$ is plotted versus trap containing particular stimuli. Parameter $\langle \lambda_1 - \lambda_2 \rangle_W$ can be used to characterize potentially ecologically relevant attractant for any tested fly strain. Possibly it can apply to any behavioral test conducted as in Ruebenbauer et al. (2008) using any insect whose biology and body size is similar to *Drosophila melanogaster*. Moreover any volatile substances that are of biological importance to the tested organism could be investigated that way.

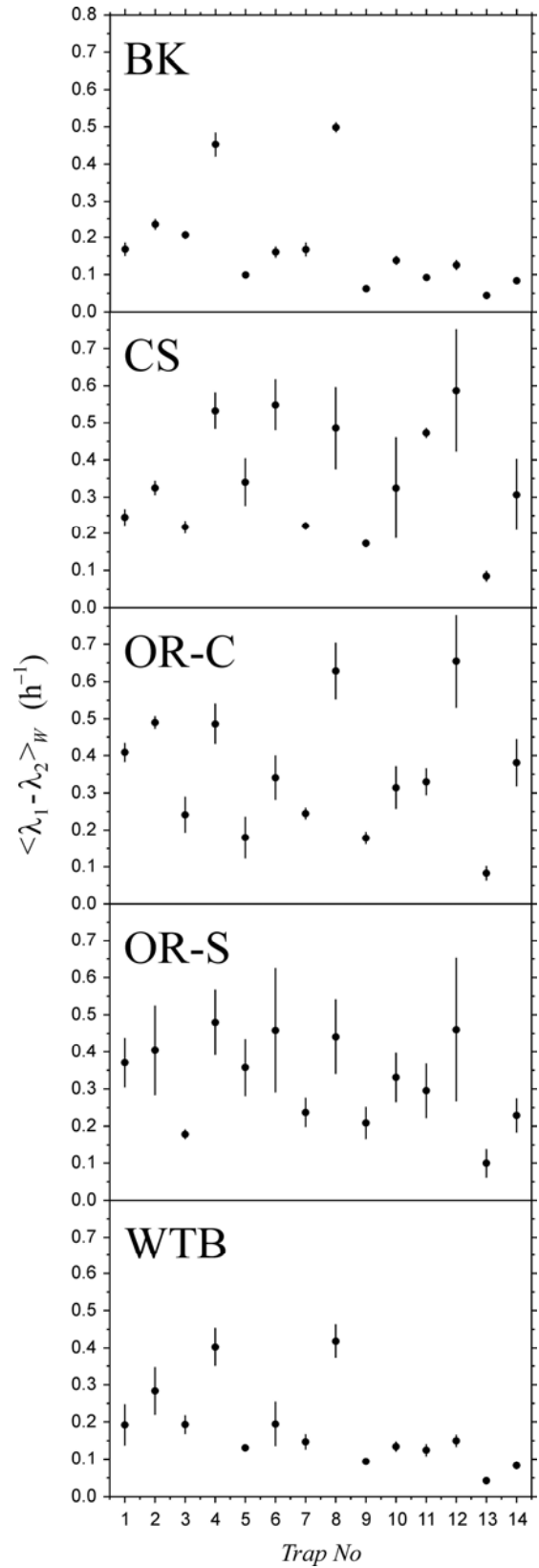

**Figure 2** Parameter $\langle \lambda_1 - \lambda_2 \rangle_W$ plotted versus *Trap No* as explained below and for various strains. Vertical bars represent effective dispersion. 1. Apple 2. Apple rotten 3. Banana 4. Banana rotten 5. Lemon 6. Lemon rotten 7. Mango 8. Mango rotten 9. Orange 10. Orange rotten 11. Strawberries 12. Strawberries rotten 13. Vinegar 14. Yeast.

For all tested strains $\langle \lambda_1 - \lambda_2 \rangle_W$ is above zero indicating that tested stimuli are attractive for the flies despite their origin. Indeed banana and other fruits have been since long time reported to be an attractive oviposition sites for *Drosophila* (Hofmann, 1985; Reed, 1938; Rkha et al., 1991; Zhu et al., 2003) and other flies (Rull and Prokopy, 2004; Staub et al., 2008). Vinegar is reported by Faucher et al. (2006) as a context and sex dependent attractant for *Drosophila melanogaster*, however there is no sex differences in vinegar attraction. Yeast was one of the laboratory diet ingredients and it is probably due to learning process that it is highly accepted by the flies (Aluja and Mangan, 2008). Moreover, it is highly explicable, that yeast serves as a good stimulus for tested flies. Since Wagners study in fourties it has been known that microbes and yeast in particular has a significant impact in *Drosophila* nutrition and ecology (Wagner, 1944; Wagner, 1949). Simply nutrition added from microbes is necessary for *Drosophila* females to mature their eggs (Begon, 1982). The parameter $\langle \lambda_1 - \lambda_2 \rangle_W$ is above 0.5 in case of all tested lines for rotten fruits, such as banana, lemon, mango and strawberries. Rotten fruits represent common odor profile rich of bacterial metabolites as acetoin (Xiao and Xu, 2007) and other organic decomposition compounds and fungi (Hasan, 2000). Decaying fruits remain the best ecologically relevant targets for *Drosophila melanogaster* in its natural environment, the place for eating, mating and finally egg laying.

In the cases of observed strong attraction one can see high effective dispersion of the attraction as well. Again, for all lines, rotten fruits are characterized by high effective dispersion. That means, traps containing less attractive stimuli, such as vinegar, are reached with higher accuracy. To detect the ideal stimuli takes the flies less time in the presented study however causes more mistakes from the flies while making their choice. The lines BK and WTB remain more similar to each other in contrast to three other ones, which supports the results obtained in Ruebenbauer et al. (2008). It seems that BK and WTB strains are more selective in their odor choice and making decision takes them more time but remains more accurate.

Differences in attracting potential of various traps are very significant. Trap potential (tested stimuli) is generally weaker function of the object type (tested flies strain), but differences are still measurable, i.e., far beyond the error.

The reverse of λ stands for average time that fly remains out of the trap, therefore the presented model might point out the optimal time for conducting such type of bioassay depending on fly strain and used stimulus. The advantage of the model is, beside defining trapping potential of the trap containing stimuli, taking into account control. Finally, model provides the method to reduce the obtained data set, offering one simple for interpretation attraction parameter instead of several hourly measurements. Simply, the parameter $\langle \lambda_1 - \lambda_2 \rangle_W$ represents the overall dynamic attraction to stimuli measured over time. Unraveling information concerning attraction towards ecologically relevant blends, possible host odors for the fruit flies and other herbivorous insects, is of significant importance for strategic decisions on local and international trade of fruit. There is little knowledge explaining the host choice behavior by modeling although understanding host plant status has been contentious for the last decades.